*Untersuchung einer nach den Euler'schen Vorschlägen (1754) gebauten Wasserturbine*

Investigation of a water turbine built according to Euler's proposals (1754)

Translated and Annotated by Sylvio R. Bistafa
August, 2021
E-mail: sbistafa@usp.br

Foreword

This is an annotated translation from German of *Untersuchung einer nach den Euler'schen Vorschlägen (1754) gebauten Wasserturbine* [Investigation of a water turbine built according to Euler's proposals (1754)] that reports the tests results of a modern (1944) prototype of the so-called Segner-Euler turbine, which was strictly constructed according to Euler's prescription as laid down in E222 -- *Théorie plus complète des machines qui sont mises en mouvement par la réaction de l'eau*. (*Mémoires de l'académie des sciences de Berlin 1756,* Vol. 10, pp. 227-295.), showing the feasibility of Euler's original proposal. A reproduction of the original paper is attached at the end of the translation.

Investigation of a water turbine built according to Euler's proposals (1754)

From Prof. Dr. J. ACKERET, E. T. H. Zurich

*Swiss construction newspaper*, volume (year): 123/124 (1944)

As is well known, Leonhard Euler made fundamentally important contributions in the field of applied mathematics and mechanics, in addition to his grandiose work on pure mathematics. Whenever he saw a possibility of a rational treatment, he seized it, and because he was not discouraged by the primitive state of the technology at the time, he found completely new connections and relationships and came up with proposals that were probably inexecutable at his time. But, eventually, became common property of technology over the course of the next two centuries. Among the numerous new ideas that were previously buried in inaccessible academy reports (but can now be published thanks to a generous donation from industry, trade and the public of Switzerland), one stands out due to its particular fruitfulness: the invention of the guide apparatus for turbines[1]. It is the result of a precise analysis of the losses in the so-called Segner water wheel. In 1750, Andreas Segner, professor in Göttingen, based on earlier proposals by Daniel Bernoulli (1738), specified the construction of a pure reaction wheel, which was carried out several times at the time, and later appeared again and again (for example, Parsons used such wheels for steam turbines in 1893, whereby he however, it has a very old forerunner in Heron of Alexandria, 120 BC). The pure reaction wheel was never able to establish itself in larger systems; but today it still ekes out a useful application in the form of the small wheels that are used as water sprinklers in gardens.

---

[1] *Théorie plus complète des machines qui sont mises en mouvement par la réaction de l'eau*. Mém. de l'acad. d. sc. de Berlin 1754 (read 13. Sept. 1753).

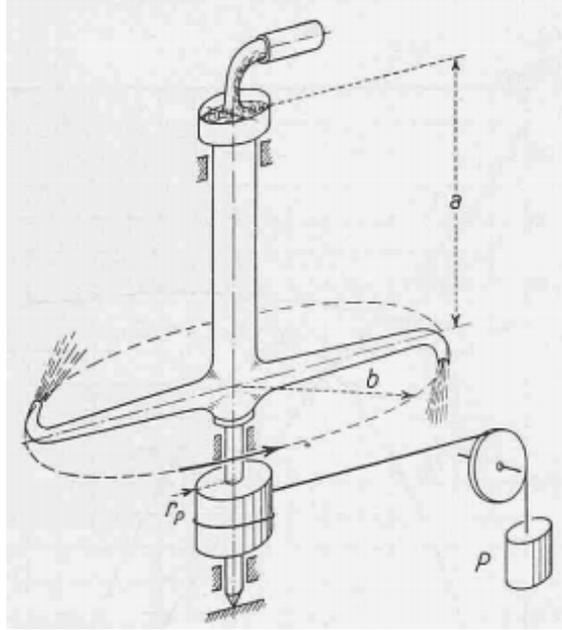
Fig. 1. Segner's reaction wheel

In order to recognize the connections in the simplest way possible, we want to use modern expressions and spelling, although Euler's old formulas contain all essential mechanical relationships in their entirety. At least, we must not forget that at that time (around 1750) the general principle of conservation of energy was not known at all, so it is a special achievement that Euler was able to deal with the question of efficiency in such a satisfactory way.

If we look at the simple diagram of the Segner wheel in Fig. 1, the torque can easily be derived from the law of angular momentum. One only has to consider that the Coriolis force do work during the rotation, which is found in the form of an increased head $H_a$ in front of the outlet opening.

Namely:

$$H_a = a + \frac{u^2}{2g}$$

where $u$ is the peripheral speed; the relative outflow velocity is calculated without friction:

$$\sqrt{2ga + u^2}$$

the absolute [velocity]:

$$\sqrt{2ga + u^2} - u$$

so this is always positive; and there is always a loss.

The torque will be:

$$2\frac{\gamma}{g} Q \left\{ \sqrt{2ga + u^2} - u \right\} b$$

$Q =$ is the flow rate per pipe and the power is:

$$L = \frac{\gamma}{g} Q \left\{ \sqrt{2ga + u^2} - u \right\} 2u$$

The efficiency is thus:

$$\eta = 2k_u \left\{ \sqrt{1 + k_u^2} - k_u \right\} 2u \quad \text{mit } k_u = \frac{u}{\sqrt{2ga}}$$

From the curve in Fig. 2 it can be seen that quite high circumferential speeds ($k_u > 1$) are necessary in order to achieve a satisfactory efficiency. Euler does not forget, however, that with increasing $k_u$ the other losses, such as pin and gear wheel friction, air resistance, etc. increase sharply, making large values of $k_u$ uneconomical.

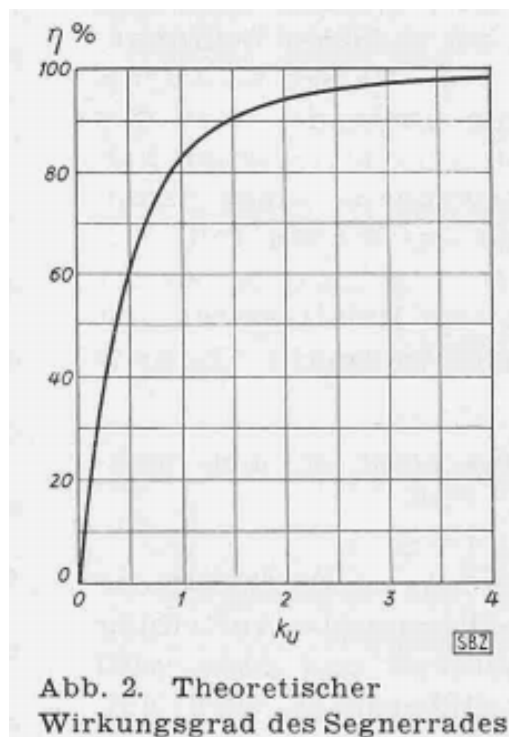

Abb. 2. Theoretischer
Wirkungsgrad des Segnerrades

Fig. 2. Theoretical efficiency of the Segner wheel

Now he [Euler] thinks up what a turbine device should look like, which in principle would be able to draw all the energy out of the water at a finite peripheral speed. As I said, the loss comes about because the water still has a finite absolute exit velocity. Euler's ingenious idea is the use of a fixed guide apparatus that feeds the water to the moving impeller at a finite tangential speed (Fig. 3).

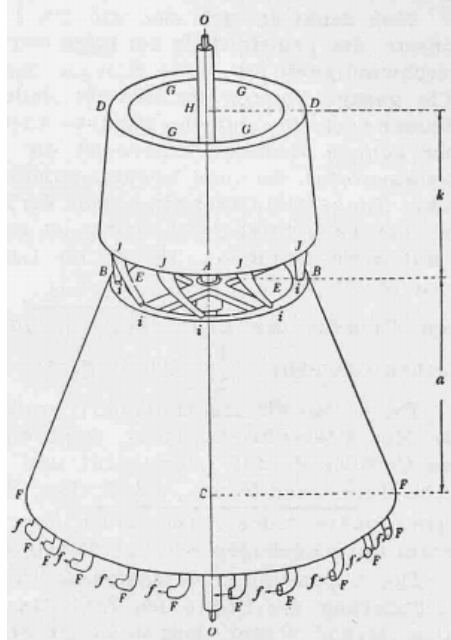

Fig. 3. Euler's proposal for a reaction turbine
with idler wheel. (According to E. Brauer and M. Winkelmann, the figure is somewhat improved compared to the Euler original)

For practical reasons, he sets the tangent of the angle of the stator at the outlet $= \frac{1}{2}$. So that: $c_{m1} = \frac{u_1}{2}$.

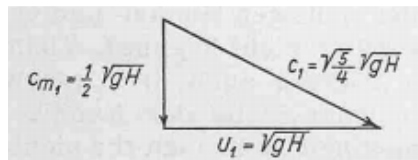

Fig. 4a. Velocity triangle at the stator outlet

The impeller now evades the tangential speed; this is thus eliminated in the relative system. Since it assumes normal entry in the impeller, then,

$$c_{u1} = u_1$$

[Here $c_{u1}$ is the tangential component of the absolute velocity of the water at the impeller inlet, $u_1$ is the tangential velocity of the impeller at the inlet.]

Since Euler assumes the impeller outlet radius to be one and a half times the inlet radius, he assumes a certain division of the head $H$ between the stator and the impeller (Fig. 3). $k = 5/8H$, $a = 3/8H$. The pressure at the gap is set equal to zero (atmospheric pressure). The mouthpieces of the impeller tubes are bent back horizontally as in Segner's wheel (Fig. 3).

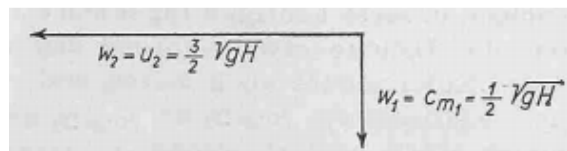

Fig. 4b. Velocity triangle at the impeller inlet

The adjustment to different amounts of water is done by changing the width of the inlet ring, whereby Euler takes care that the radii are not too large.

With a frictionless flow one obtains for this arrangement (Fig. 4):

$$u_1 c_{u1} = u_1^2 = gH \text{ (Euler's turbine equation)}$$

$$c_{u1} = u_1 = \sqrt{gH}$$

$$c_{m1} = 0{,}5\sqrt{gH} = w_1$$

[Here $w_1$ is the relative velocity at the impeller inlet.]

$$c_1 = \sqrt{1{,}25}\sqrt{gH} = \sqrt{5/8}\sqrt{2gH}$$

[Here $c_1$ is the absolute velocity at the impeller inlet.]

A relative acceleration takes place in the impeller due to the head at the impeller and the centrifugal force:

$$\frac{w_2^2}{2g} = \frac{w_1^2}{2g} + \frac{3}{8}H + \frac{u_2^2 - u_1^2}{2g}$$

Then, by making the due substitutions $u_2 = 1{,}5 u_1$:

[Here $u_2$ is the tangential velocity of the impeller at the outlet.]

$$\frac{w_2^2}{2g} = 0{,}125H + \frac{3}{8}H + 1{,}25\frac{H}{2} = \frac{9}{8}H$$

$$w_2 = \sqrt{\frac{9}{4}gH} = \frac{3}{2}u_1 = u_2$$

[Here $w_2$ is the relative velocity at the impeller inlet.]

i.e., the absolute exit velocity is now zero, and the theoretical efficiency= 1.

It seemed interesting to me to have a look at the efficiency of the Euler-designed turbine. The *Escher Wyss* Company, which was just celebrating its hundredth year of turbine construction, very gratefully prepared to take over the construction of the machine; a sign that even today the sense of history has not died out. It is designed for a head of 1 m, a water discharge of 19.7 l/sec and a rotation speed of 300/min (Fig. 5). The mean diameter at the inlet of the impeller is 200 mm; at the exit 300 mm. Although Euler points out that the water flow decreases due to friction, and the turbine is to be expected for an actually smaller head, we have not made any changes to the theoretical angles and opening cross-sections, as Euler naturally still lacks precise information about this.

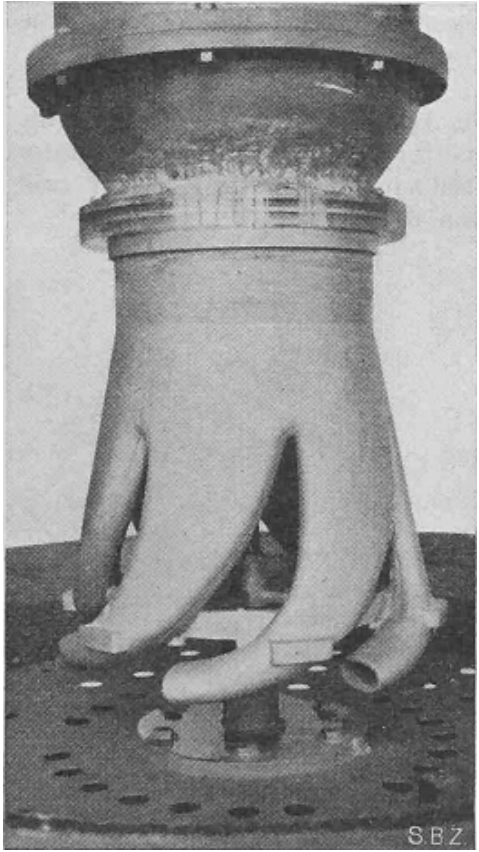

Fig. 5. Stator of the test turbine

Soon after commissioning, it turned out to be necessary to seal the gap to the outside, as the wheel lost a lot of water. We attached a simple gap seal. One can trust contemporary mechanical engineering, which at least managed to create large dewatering machines and pumping stations, that it might have helped itself in some similar way. The water is supplied evenly through special distribution pipes. The head is calculated up to the lower edge of the impeller (Fig. 6); the amount of water is determined by a VDI standard screen that we have calibrated directly (by water weighing). A cord drum served as a brake; the very low ball bearing friction has been subtracted. The results are shown in Fig. 7. As expected, the speed for the best efficiency turns out to be significantly lower than according to the calculation; this expresses the complete neglect of friction. The torque curve is almost straight, so there is a very favorable starting torque that would undoubtedly have been very advantageous for the work machines at that time. The amount of water changes only slightly, the efficiency parabolic. In view of the complete neglect of friction in the design and the very low output of only 0.15 PS, the maximum value of 0.71 must be described as quite satisfactory even for today's views, as modern Francis wheels, the *Thoma* have similarly small outputs and at the only investigated head of 1 m, yielded values of 0.78 to 0.82.

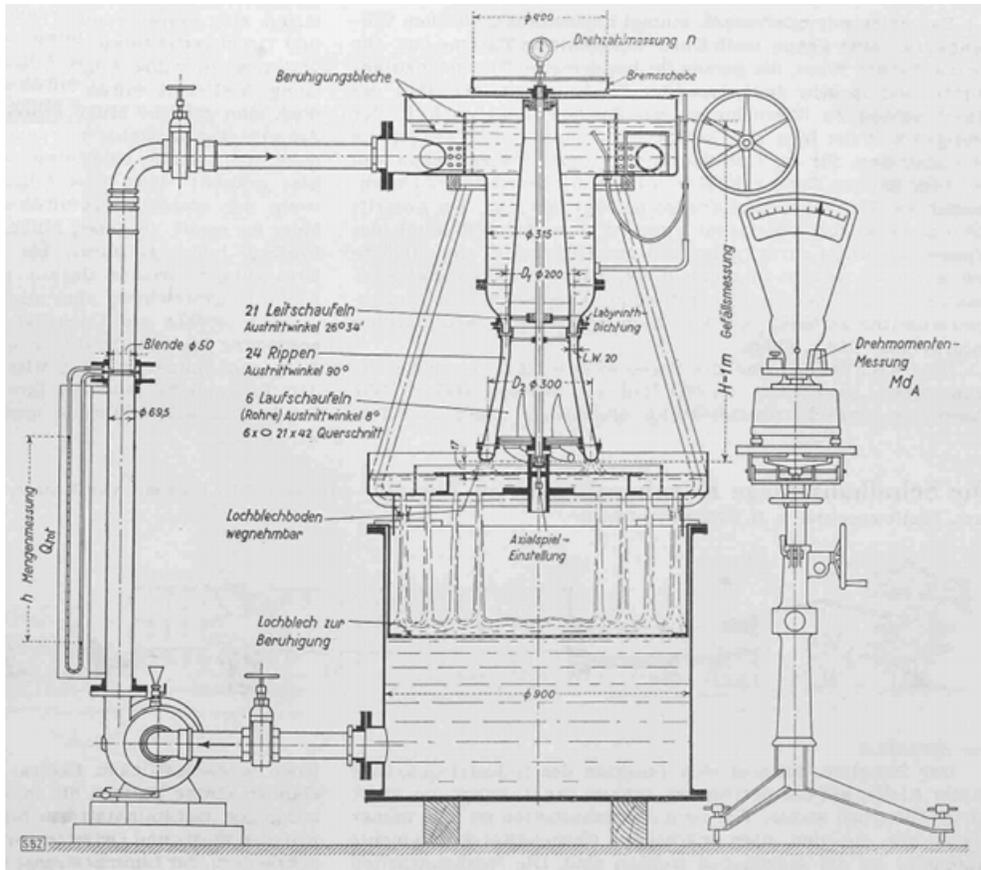

Fig. 6. Euler turbine, experimental set-up, - scale of the drawing 1: 20

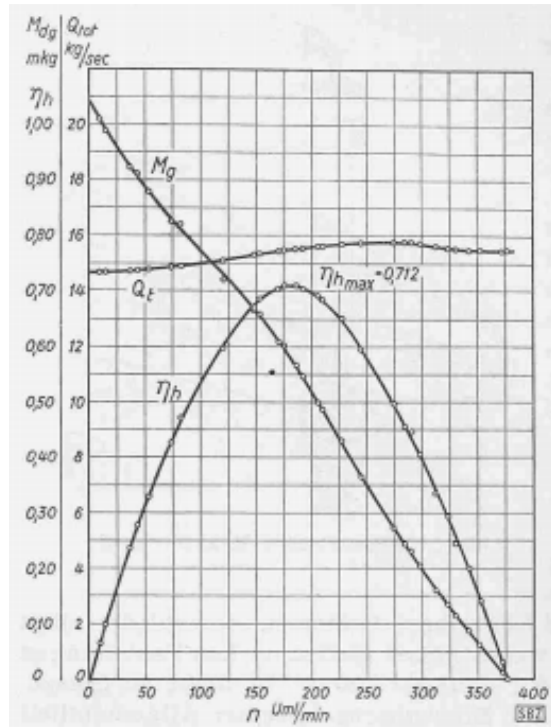

Fig. 7. Test results

Euler was not satisfied with finding the correct angle and cross-sectional relationships. Rather, through an extensive analysis, he also examined the speed and pressure conditions at every point in the ducts, in full generality, i.e. also for non-stationary conditions. Perhaps it would take a certain amount of effort for many of our engineers today to specify the pressures, for example when the machine is starting up. Euler does this too, and in doing so, he makes an important discovery in that he clearly recognizes (1754!) that the flow must be severely disturbed when the absolute pressure drops to zero. The cavitation problem has thus been shown; but it took a full one hundred and fifty years to come up with it again through all sorts of bad experiences; today cavitation research is an extensive science.

I don't want to forget the Escher Wyss company, especially Ing. C. *Keller* and Dir. Ing. Dr. H. *Gygi* for their kindness, to thank my colleagues Dipl. Ing. E. *Mühlemann* and technician J. *Egli* for their cooperation in the design and the execution of the tests.

Untersuchung einer nach den Euler'schen Vorschlägen (1754) gebauten Wasserturbine

Von Prof. Dr. J. ACKERET, E. T. H. Zürich

Schweizerische Bauzeitung, Band (Jahr): 123/124 (1944)

Leonhard Euler hat bekanntlich ausser seinen grandiosen Arbeiten zur reinen Mathematik auch auf dem Gebiete der angewandten Mathematik und Mechanik grundlegend wichtige Beiträge geliefert. Wo immer er eine Möglichkeit einer rationellen Behandlung sah,griff er zu, und dadurch, das ser sich nicht durch den primitive Zustand der damaligen Technologie entmutigen liess, fan der ganz neue Zusammenhängeunde Beziehungen und kam zu Vorschlägen, die zu seiner Zeit wohl unausfuhrbar waren, im Laufe der nächsten zwei Jahrhunderte aber schliesslich zum Allgemeingut der Technik wurden. Unter den zahlreichen neuen Ideen, die bisher in wenig zugänglichen Akademieberichten vergraben lagen (nunmehr aber dank einer grosszügigen Spende von Industrie, Handel und öffentlic der Scweiz gesammelt herausgegeben warden können), ragt eine durch ihre besondere Fruchtbarkeit hervor: die Erfindung des Leitapparates für Turbinen[2]). Sie ist das Ergebnis einer genauen Analyse der Verluste im sog. Segner'schen Wasserrad. Andreas Segner, Professor in Göttingen, gab 1750 in Anlehnung an frühere Vorschläge von Daniel Bernoulli (1738) die Konstruktion eines reinen Reaktionsrades an, das seinerzeit mehrfach ausgeführt wurde und spatter immer wieder auftauchte (beispielsweise hat Parsons 1893 solche Räder für Dampfturbinen benützt, wobei er allerdings einen sehr alten Vorläufer in Heron von Alexandrien, 120 v. Chr., hat). Durchsetzen konnte sich das reine Reaktionsrad in grösseren Anlagen nie; es fristet aber heute noch ein gemütliches Leben in Form der kleinen Rädchen, die als Wassersprenger in Gärten verwendet werden.

Wir wollen, um die Zusammenhänge auf möglichst einfache Weise zu erkennen, uns der modernen Ausdrucksweise und Schreibart bedienen, obwohl die alten Formeln von Euler alle wesentlichen mechanischen Beziehungen vollständig enthalten. Wir müssen immerhin nicht vergessen, dass damals (um 1750) der Energiesatz in seiner Allgemeinheit überhaupt nicht bekannt war, es also eine besondere Leistung darstellt, dass Euler gerade die Frage des Wirkungsgrades in so befriedigender Weise erledigen konnte.

Betrachten wir das einfache Schema des Segnerrades Abb. 1, so ist das Drehmoment aus dem Drehimpulssatz leicht abzuleiten. Man muss nur bedenken, dass die Corioliskräfte bei der Drehung eine Arbeit leisten, die sich in Form erhöhten Gefälles $H_a$ vor der Ausströmöffnung findet.

---

[2] Théorie plus complète des machines qui sont mises en mouvement par la réaction de l'eau. Mém. de l'acad. d. se. de Berlin 1754 (gelesen 13. Sept. 1753).

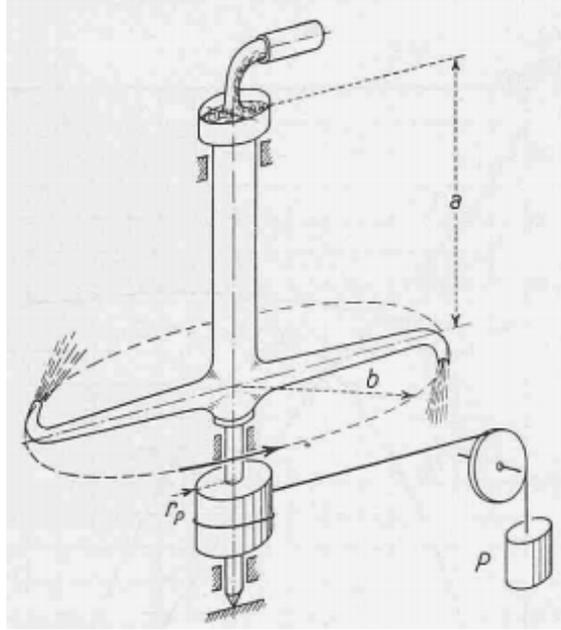

Abb. 1. Segner'sches Reaktionsrad

Nämlich :

$$H_a = a + \frac{u^2}{2g}$$

wo $u$ die Umfangsgeschwindigkeit ist; die relative Ausströmgeschwindigkeit ist, reibungsfrei gerechnet:

$$\sqrt{2ga + u^2}$$

die absolute:

$$\sqrt{2ga + u^2} - u$$

diese ist also immer positiv; es ist stets ein Verlust vorhanden.

Das Drehmoment wird:

$$2\frac{\gamma}{g} Q \left\{ \sqrt{2ga + u^2} - u \right\} b$$

$Q =$ Wassermenge pro Rohr die Leistung:

$$L = \frac{\gamma}{g} Q \left\{ \sqrt{2ga + u^2} - u \right\} 2u$$

Der Wirkungsgrad ist somit:

$$\eta = 2k_u \left\{ \sqrt{1 + k_u{}^2} - k_u \right\} 2u \quad \text{mit } k_u = \frac{u}{\sqrt{2ga}}$$

Aus der Kurve Abb. 2 ersieht man, dass schon ziemlich hohe Umfangsgeschwindigkeiten $(k_u > 1)$ nötig sind, um eine befriedigende Ausnützung zu bekommen. Euler vergisst aber nicht, dass mit wachsendem ku die sonstigen Verluste, wie Zapfen- und Zahnräderreibung, Luftwiderstand usw. stark anwachsen und grosse $k_u$ unwirtschaftlich machen.

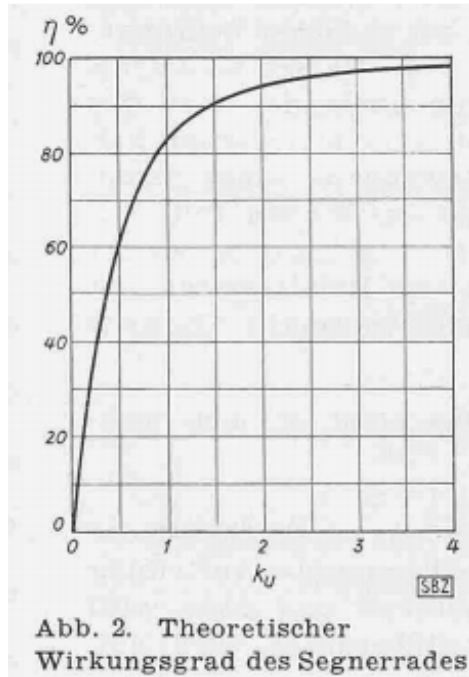

Abb. 2. Theoretischer Wirkungsgrad des Segnerrades

Nun denkt er sich aus, wie ein Turbinenapparat aussehen musste, der prinzipiell in der Lage wäre, bei *endlicher* Umfangsgeschwindigkeit die *ganze* Energie aus dem Wasser zu ziehen. Wie gesagt kommt der Verlust dadurch zustande, dass das Wasser noch eine endliche absolute Austrittsgeschwindigkeit hat. Der geniale Gedanke Eulers ist die Anwendung eines festen *Leitapparates*, der dem bewegten *Laufrad* das Wasser mit endlicher Tangential-Geschwindigkeit zuführt. Das Laufrad weicht nun der Tangential-Geschwindigkeit aus; im Relativsystem ist somit diese eliminiert. Da er im Laufrad normalen Eintritt voraussetzt, ist

$$c_{u1} = u_1$$

Den Tangens des Leitradaustrittswinkels setzt er aus praktischen Gründen $= \frac{1}{2}$. Damit wird: $c_{m1} = \frac{u_1}{2}$.

Da Euler als Laufradaustrittsradius das anderthalbfache des Eintrittsradius annimmt, folgt eine bestimmte Aufteilung des Gefälles $H$ auf Leitapparat und Laufrad und zwar wird (Abb. 3). $k = 5/8H$, $a = 3/8H$. Der Spaltüberdruck wird Null (Atmosphärendruck). Die Enden der Laufradröhren sind horizontal zurückgebogen wie bei Segner (Abb. 3).

Die Anpassung an verschiedene Wassermengen erfolgt durch Veränderung der Breite des Eintrittsringes, wobei Euler sorgfältig darauf achtet, dass nicht zu grosse Unterschiede in den Radien auftreten dürfen.

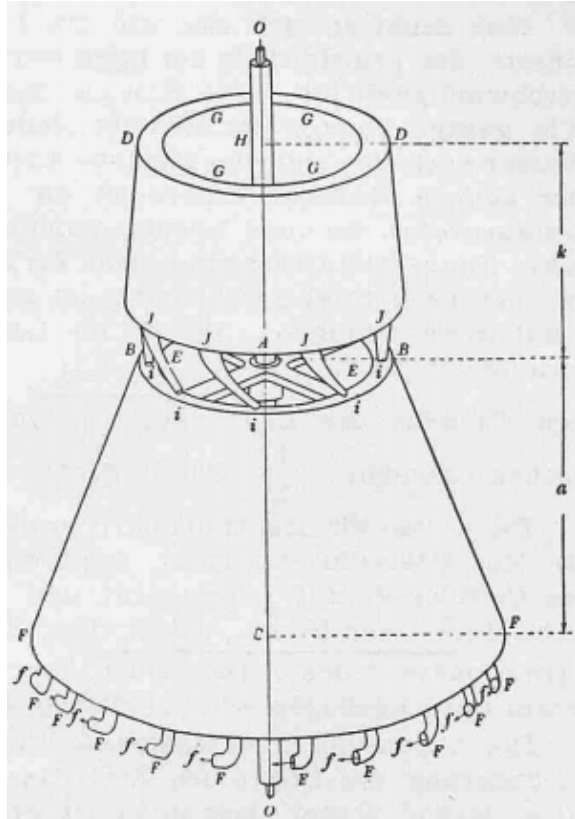

Abb. 3 (rechts). Vorschlag von Euler für eine Reaktionsturbine
mit Leitrad. (Die Figur ist nach E. Brauer und M. Winkelmann gegenüber dem Euler'schen Original etwas verbessert)

Bei reibungsfreier Strömung erhält man für diese Anordnung (Abb. 4):

$$u_1 c_{u1} = u_1^2 = gH \;(\text{Euler'sche Turbinengleichung})$$

$$c_{u1} = u_1 = \sqrt{gH}$$

$$c_{m1} = 0{,}5\sqrt{gH} = w_1$$

$$c_1 = \sqrt{1{,}25}\sqrt{gH} = \sqrt{5/8}\sqrt{2gH}$$

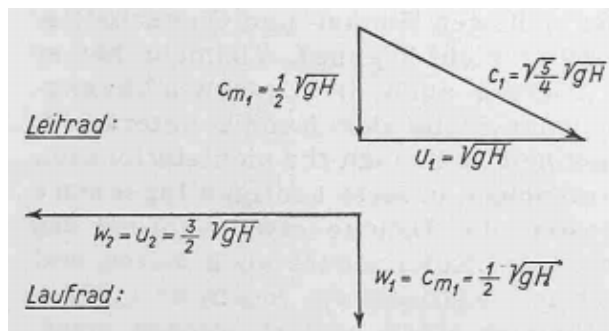

Abb. 4. Geschwindigkeitsdreiecke der Eulerturbine

Im Laufrad erfolgt eine relative Beschleunigung durch das Laufradgefälle und die Zentrifugalkraft:

$$\frac{w_2^2}{2g} = \frac{w_1^2}{2g} + \frac{3}{8}H + \frac{u_2^2 - u_1^2}{2g}$$

Setzen wir ein, so ergibt sich mit $u_2 = 1{,}5 u_1$:

$$\frac{w_2^2}{2g} = 0{,}125 H + \frac{3}{8}H + 1{,}25\frac{H}{2} = \frac{9}{8}H$$

$$w_2 = \sqrt{\frac{9}{4}gH} = \frac{3}{2}u_1 = u_2$$

d. h., die absolute Austrittsgeschwindigkeit ist jetzt null, der theoretische Wirkungsgrad $= 1$.

Es schien mir interessant, einmal nachzusehen, welchen Wirkungsgrad sine genau nach Euler konstruierte Turbine hat. Firma *Escher Wyss*, die gerade ihr hundertstes Turbinenbaujahr feierte, war in sehr dankenswerter Weise bereit, den Bau des Maschinchens zu übernehmen; ein Zeichen, dass auch in der heutigen Zeit der Sinn für Geschichte nicht ausgestorben ist. Es ist konstruiert für ein Gefälle von 1 m, eine Wassermenge von 19,7 l/sec und die Drehzahl 300/min (Abb. 5). Der mittlere Durchmesser am Eintritt des Laufrades beträgt 200 mm; am Austritt 300 mm. Obwohl Euler darauf hinweist, dass durch Reibung der Wasserdurchfluss zurückgeht, und die Turbine für ein tatsächlich kleineres Gefälle zu rechnen ist, haben wir keinerlei Aenderungen gegenüber den theoretischen Winkeln und Oeffnungs-Querschnitten gemacht, da bei Euler präzise Angaben darüber naturgemäss noch fehlen.

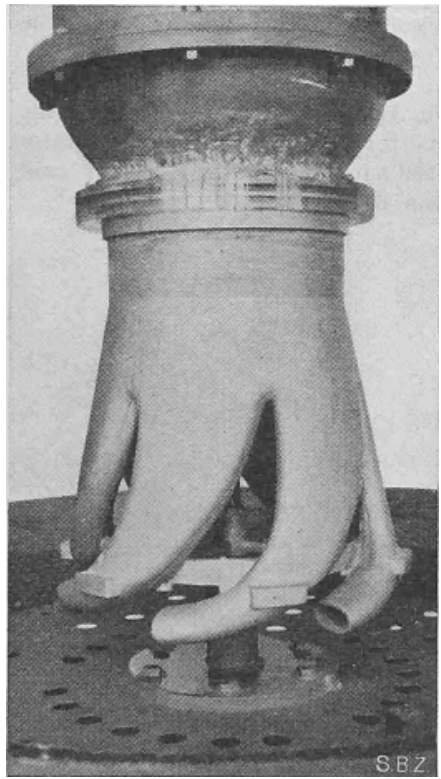

Abb. 5. Laufrad der Versuchsturbine

Bald nach Inbetriebnahme erwies es sich als nötig, den Spalt nach aussen zu dichten, da das Rad viel Wasser verlor. Wir haben eine einfache Spaltdichtung angebracht. Man darf dem zeitgenössischen Maschinenbau, der immerhin schon grosse Wasserhaltungs – Maschinen und Pumpwerke zustande brachte, zutrauen, dass er sich gegebenenfalls auf irgend eine ähnliche Weise geholfen hätte. Durch besondere Verteilrohre wird das Wasser gleichmässig zugeführt. Das Gefälle wird bis Unterkante Laufrad gerechnet (Abb. 6); die Wassermenge wird durch eine VDI-Norm- blende, die wir direkt (durch Wasserwägung) geeicht haben bestimmt. Als Bremse diente eine Schnur-Trommel; die sehr geringe Kugellagerreibung wurde abgezogen. Die Ergebnisse sind in Abb. 7 dargestellt. Die Drehzahl für den besten Wirkungsgrad erweist sich erwartungsgemäss als ebendie bedeutend niedriger als nach Rechnung; hierin äussert sich eben die vollständige Vernachlässigung der Reibung. Der Momenten-Verlauf ist annähernd geradlinig, es ist also ein recht günstiges Anfahrmoment da, das für den damaligen Stand der Arbeits-Maschinen zweifellos sehr vorteilhaft gewesen wäre. Die Wassermenge ändert sich nur wenig, der Wirkungsgrad parabolisch. Der Höchstwert von 0,71 muss in Anbetracht der völligen Reibungsvernachlässigung beim Entwurf und der sehr geringen Leistung von nur 0,15 PS auch für heutige Anschauungen als recht recht befriedigend bezeichnet werden, haben doch modern Francisräder, die *Thoma* für ähnlich kleine Leistungen und 1 m Gefälle gebaut und untersucht hat, nur Werte von 0,78 bis 0,82 ergeben.

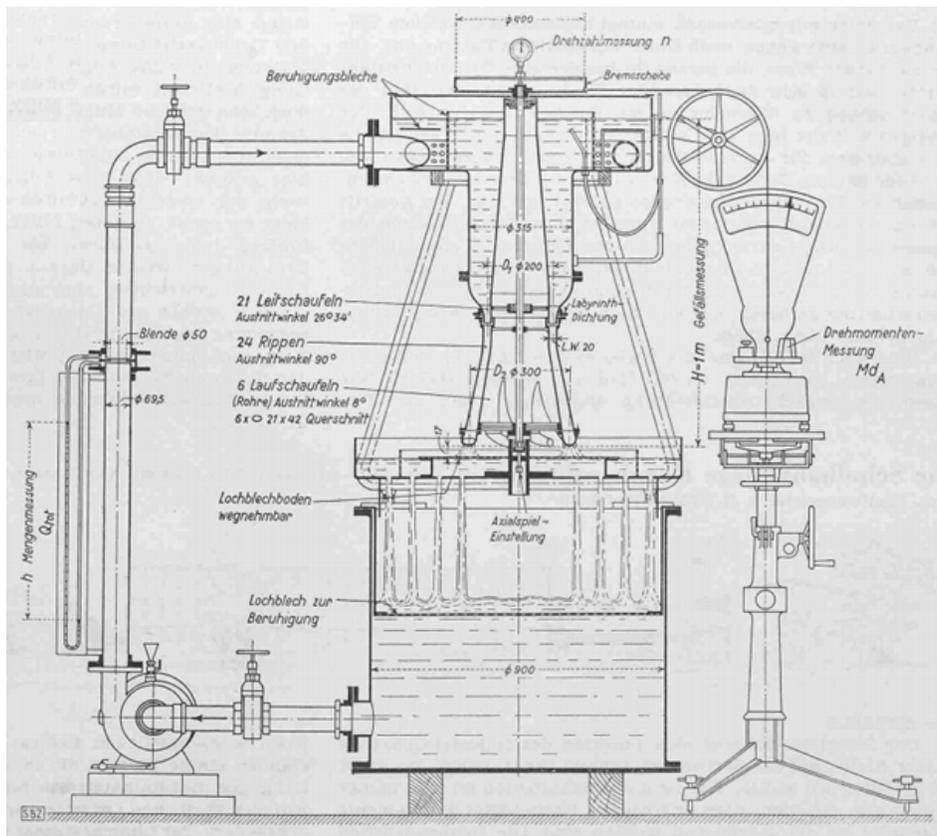

Abb. 6.   Eulerturbine, Versuchsanordnung, — Masstab der Zeichnung 1: 20

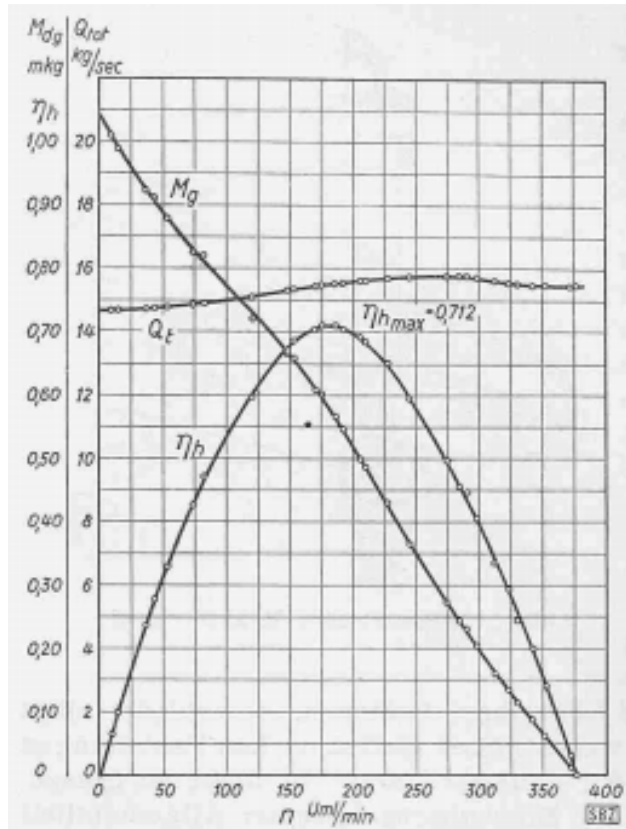

Abb. 7.  Versuchsergebnisse

Mit der Auffindung der richtigen Winkel- und Querschnitts- Zusammenhänge hat sich Euler nicht begnügt. Vielmehr hat er durch eine umfangreiche Analyse auch die Geschwindigkeitsund und Druckverhältnisse an jeder Stelle der Kanäle untersucht, und zwar in voller Allgemeinheit, d. h. auch für nichtstationären Gang. Vielleicht würde es manchem unserer heutigen Ingenieure doch eine gewisse Mühe kosten, die Drücke etwa während des Anlaufs der Maschine anzugeben. Euler leistet auch dieses, und dabei gelingt ihm eine wichtige Entdeckung, indem er (1754!) klar erkennt, dass die Strömung stark gestört werden muss, wenn der absolute Druck auf Null sinkt. Das Kavitationsproblem ist somit aufgezeigt worden; aber es hat volle hundertfünfzig Jahre gedauert, bis man durch allerhand schlechte Erfahrungen wieder darauf gekommen ist; heute bildet die Kavitationsforschung eine umfangreiche Wissenschaft.